\documentclass[11pt,a4paper]{llncs}
\usepackage{amsmath}
\usepackage{amssymb}
\usepackage{graphicx}
\usepackage{marvosym}
\usepackage{url}
\usepackage{fancyhdr}

\usepackage{geometry}
\newcommand{\keywords}[1]{\par\addvspace\baselineskip
\noindent\keywordname\enspace\ignorespaces#1}

\fancypagestyle{firstpage}{\fancyhf{}
}

\begin{document}

%\mainmatter  % start of an individual contribution

\title{\LARGE{Lightweight Public Key Encryption in Post-Quantum Computing Era}}

%\titlerunning{Lecture Notes in Computer Science: Authors' Instructions}

\author{\large{Peter Hillmann}}
\institute{\large{University of the Bundeswehr Munich,\\ Department of Computer Science,\\ Werner-Heisenberg-Weg 39, 85577 Neubiberg, Germany}} %Institute for Applied Computer Science,

%\author{Alfred Hofmann%
%\thanks{Please note that the LNCS Editorial assumes that all authors have used
%the western naming convention, with given names preceding surnames. This determines
%the structure of the names in the running heads and the author index.}%
%\and Ursula Barth\and Ingrid Haas\and Frank Holzwarth\and\\
%Anna Kramer\and Leonie Kunz\and Christine Rei\ss\and\\
%Nicole Sator\and Erika Siebert-Cole\and Peter Stra\ss er}

%\institute{Springer-Verlag, Computer Science Editorial,\\
%Tiergartenstr. 17, 69121 Heidelberg, Germany\\
%\url{http://www.springer.com/lncs}}


%\toctitle{Lecture Notes in Computer Science}
%\tocauthor{Authors' Instructions}

\maketitle

\thispagestyle{firstpage}

\begin{abstract}
Confidentiality in our digital world is based on the security of cryptographic algorithms.
These are usually executed transparently in the background, with people often relying on them without further knowledge.
In the course of technological progress with quantum computers, the protective function of common encryption algorithms is threatened.
This particularly affects public-key methods such as RSA and DH based on discrete logarithms and prime factorization.
Our concept describes the transformation of a classical asymmetric encryption method to a modern complexity class.
Thereby the approach of Cramer-Shoup is put on the new basis of elliptic curves.
The system is provable cryptographically strong, especially against adaptive chosen-ciphertext attacks. %chosen-plaintext and
In addition, the new method features small key lengths, making it suitable for Internet-of-Things.
It represents an intermediate step towards an encryption scheme based on isogeny elliptic curves.
This approach shows a way to a secure encryption scheme for the post-quantum computing era.
\end{abstract}

%\begin{abstract}
%The abstract should summarize the contents of the paper and should
%contain at least 70 and at most 150 words. It should be written using the
%\emph{abstract} environment.
%\keywords{We would like to encourage you to list your keywords within
%the abstract section}
%\end{abstract}

\section{Introduction}
More than 50 \% of all internet traffic use the protocol combination of RSA with \textit{Optimal Asymmetric Encryption Padding}~(OAEP), i. e. in https. % and ftps.
Also the financial market relies on RSA for online-banking, even if RSA is only cryptographic weak. % compared to CS.
Attacks on \textit{Secure Sockets Layer} (SSL/TLS) following \textit{Public-Key Cryptography Standards} (PKCS) \#1 v1.5 show the weakness of the cryptosystem RSA.
For example, the approaches of \textit{Bleichenbacher}~\cite{Bleichenbacher:1998:CCA:646763.706320} could reveal the content of encrypted messages.
In order to prevent such \textit{Adaptive Chosen Ciphertext Attacks} (CCA2), it is necessary to use an encryption or encoding scheme that limits ciphertext malleability. %verformbar
To address this problem, RSA is combined with the coding scheme OAEP~\cite{Bellare1994}.
It is standardized in the updated PKCS\#1 v2 (RFC 2437, RFC 8017).
%To address this problem, the coding scheme \textit{Optimal Asymmetric Encryption Padding} (OAEP) was standardized in PKCS#1 v2 (RFC 2437-->8017), a combination of RSA and OAEP.
%ECDHE is a standard part of the transport security layer (TLS) underlying the secure hyper text transfer protocol (https). As of 2017, more than 50% of all internet traffic uses this protocol.
%Computation in supersingular isogeny graphs; Andrew V. Sutherland; MIT; 2018
%
%
The security of OAEP has been proven secure in the random oracle model~\cite{Fujisaki2004}.
This model is typically used when the proof cannot be carried out using weaker assumptions compared to the \textit{Standard Model of Cryptography} (SMC).
The SMC uses only complexity assumptions for the verification.
A growing body of evidence claims the insecurity of this approach~\cite{Canetti:2004:ROM:1008731.1008734}.
Even the improved scheme OAEP+ is only proved secure against non-adaptive \textit{Chosen Ciphertext Attacks} (CCA) in general. %ciphertext attack (IND-CCA1). % in SMC
The security is still indistinguishability under CCA2 in the SMC~\cite{Paillier2006,Brown07whathashes}. %
%Beside this, the combination of RSA with OAEP is in relation to security level indistinguishability under chosen plaintext attack (IND-CPA).
%, but not chosen non-adaptive ciphertext attack (IND-CCA1), and indistinguishability under adaptive chosen ciphertext attack (IND-CCA2).
%An improved scheme (called OAEP+) is only IND-CCA1 secure.
%More recent work has shown that in the standard model it is impossible to prove the IND-CCA2 security of RSA-OAEP under the assumed hardness of the RSA problem.
Furthermore, vulnerabilities still exist with slight variations like \textit{Return Of Bleichenbacher's Oracle Threat} since 20 years~\cite{217494}.
%Done: https://www.golem.de/news/robot-angriff-19-jahre-alter-angriff-auf-tls-funktioniert-immer-noch-1712-131607.html
The history shows that the current improvements have not solved the problem fundamentally by modified attacks~\cite{Manger2001,Ronen2018}.
%Quelle: James Manger: A Chosen Ciphertext Attack on RSA Optimal Asymmetric Encryption Padding (OAEP) as Standardized in PKCS #1 v2.0. In: CRYPTO 2001 (= Lecture Notes in Computer Science). vol. 2139. Springer, 2001, S. 260–274

However, the combination of RSA with OAEP was favored instead of using a different encryption system with inherent strength against such attacks. % like \textit{Cramer-Shoup}~\cite{Cramer2003}. %CCA.
Nevertheless, the confidentiality of crypto systems is threatened by the rising quantum computing era.
This also has a significant impact on the trustworthiness of all current blockchain applications~\cite{Heiland2022} due to the public-key procedures used.
More complex mathematical problems need to be identified for cryptography as large number factorization is developed.
%More complex mathematical problems have to be identified for cryptography due to the development on large number factorization.
The algorithms of \textit{Shore} and \textit{Grover} allow fast factorization and search.
This particularly affects public-key methods based on discrete logarithms and prime factorization such as RSA and DH.
%computers, the protective function of common encryption algorithms is threatened. This particularly affects public-key methods such as RSA and DH based on discrete logarithms and prime factorization.
%, we faces another security challenge with the rise of quantum computer.
To address this problem fundamentally, we enhance a public-key encryption system for the \textit{post-quantum cryptographic}~(PQC) age.
%The direction of elliptic curve cryptography has yielded algorithms based on the discrete logarithm problem. % by Neal Koblitz and Victor Miller. %1985
%It provides smaller key sizes and faster operations for approximately equivalent estimated security. 
The direction of \textit{Elliptic Curve Cryptography}~(ECC) has yielded new algorithms, which provides increased security. % for approximately equivalent estimated security.
%To address this problem fundamentally, we develop on a classical asymmetric crypto system to develop it further for the new cryptographic age.
Our base is the \textit{Cramer-Shoup} crypto system, which is mathematical proven secure against CCA2 in SMC~\cite{Cramer2003}. %under standard intractability assumptions
This security definition is currently the strongest confidentiality proof known for a public-key crypto system.
This prevents such attacks like on RSA-OEAP from the beginning.
Our contribution focuses on increasing security while keeping the keys small.
In this paper, we highlight the requirements on modern crypto systems with focus on \textit{Internet of things} (IoT).
In our concept, we develop the \textit{Cramer-Shoup}~(CS) crypto system further to be resistant against quantum computing possibilities.
Therefore, we adapt CS to the mathematical base of \textit{Elliptic Curves} (EC). % like from DH to EC-DH or RSA to ECC
%To the best of our knowledge, this has not yet been done.
The main advantages are shorter keys, faster operations and increased security compared to the algorithms based on classical discrete logarithms.
This is especially desired in lightweight cryptographic for mobile and wireless applications as in the IoT environments~\cite{Hillmann2019}.
In addition, we provide an simple and detailed description for comprehensible implementation, also with regard to further research.
%The same methodology has been done like from DH to EC-DH or RSA to ECC.
Based on this approach, we will extend our solution to supersingular isogeny EC or graphs for higher protection class in a future step.
This new mathematical construct is promising to be resistant to attacks via quantum computers in 21st century.
%Our contribution focuses on increasing security while keeping the keys small.
Beside this, we give an overview about the development of public-key schemes and provide a performance comparison. % and provide a collection of requirements for modern crypto systems.

%In a next step, we extend our system to supersingular isogeny ellipctic curves to further increase the security level specific for qunatum computer resistans.

The structure of this paper is as follows: Section 2 describes a typical security scenario and lists the
requirements for modern crypto systems. In Section 3, we provide an overview of the current state of the art
with focus on EC. The main part in Section 4 describes our concept of a public-key crypto system.
Then, we show the correctness of the our approach and evaluate the performance in comparison to other approaches in
Section 5. Subsequently, we proof the fundamental security properties of the presented system in Section 6. A discussion on security in the post-quantum era is elaborated in Section 7. The last section summarizes our work and provides an outlook.

\section{Scenario and Requirements}
\begin{figure}[htbp]
	\centering
	\includegraphics[width=0.9\linewidth]{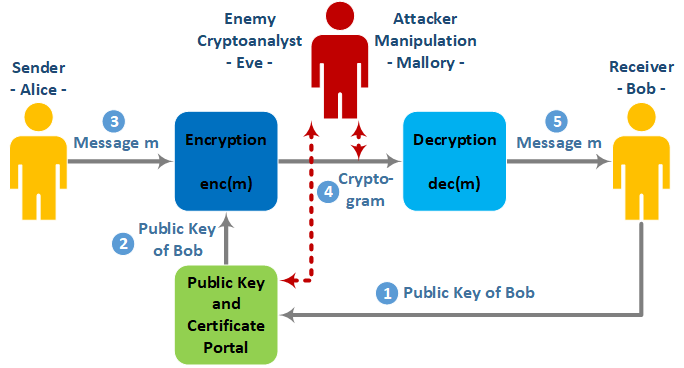}
	\caption[]{General secrecy system scenario~\cite{Shannon1946}.}
	\label{fig:scenario}
\end{figure}

Our approach is based on the following common scenario for public-key encryption, see Figure~\ref{fig:scenario}.
%\todo[inline]{selbst malen}
%scenario_general_secrecy_system_shannon.png
%Communication Theory of Secrecy Systems ?
%By C. E. S HANNON: A Mathematical Theory of Cryptogra-
%phy”dated Sept.1, 1946
A sender wishes to send a confidential message to a particular recipient.
For that purpose, the recipient has shared the public part of his asymmetric key on a free portal on the Internet (1).
This is preferably included in a cryptographic certificate.
The sender uses this public key (2) to encrypt the message (3).
The encrypted message is then sent to the receiving party (4).
The recipient can decrypt the message and process it based on knowledge of the corresponding private key (5).
An omnipotent attacker can access both the public part of the key and the encrypted message.
Therefore, the encryption method must provide cryptographically strong protection.

%Why not choose a asymmetric crypto system, which is mathematical provable secure against adaptive chosen ciphertext attack using standard cryptographic assumptions from start (IND-CCA2).
%Though it was well known that many widely used cryptosystems were insecure against such an attacker, for many years system designers considered the attack to be impractical and of largely theoretical interest.
%This began to change during the late 1990s, particularly when Daniel Bleichenbacher demonstrated a practical adaptive chosen ciphertext attack against SSL servers using a form of RSA encryption. [1]

The following requirements are mandatory for modern crypto systems:
\begin{itemize}
	\item Tiny keys for fast transmission %, especially for Internet-of-Things.
	\item Forward secrecy and reusable keys
	\item Integrated message validation
	\item No Malleability~\cite{Bellare1998}
	\item Provable strong security against adaptive CPA and CCA2, highest possible security
	\item Additionally in software engineering~\cite{Bernstein2018}:
	\begin{itemize}
	\item No data flow from secrets to array indices % and branch conditions
	\item No data flow from secrets to branch conditions
	\item No padding oracles
	\item Centralizing randomness
	\item Avoiding unnecessary randomness with focus on audited deterministic functions
	%\item Eliminate low-security options
	\end{itemize}
\end{itemize}
%New requirements for crypto software engineering to avoid real-world crypto disasters:
%No data flow from secrets to array indices. Stops, e.g., 2016 CacheBleed attack.
%No data flow from secrets to branch conditions. Stops, e.g., 2018 RSA key-generation attack by Aldaya–Garcıa–Tapia–Brumley.
%No padding oracles. Stops, e.g., 2017 ROBOT attack.
%But wait, there’s more:
%Centralizing randomness: system has one central audited fast PRNG. Stops, e.g., Juniper fiasco discovered in 2015.
%Avoiding unnecessary randomness: use audited deterministic functions. Stops, e.g., 2017 ROCA attack
%Eliminate low-security options. Stops, e.g., 2015 Logjam attack.
\section{History and Related Work}
\begin{figure}[htbp]
	\centerline{\includegraphics[width=0.8\textwidth]{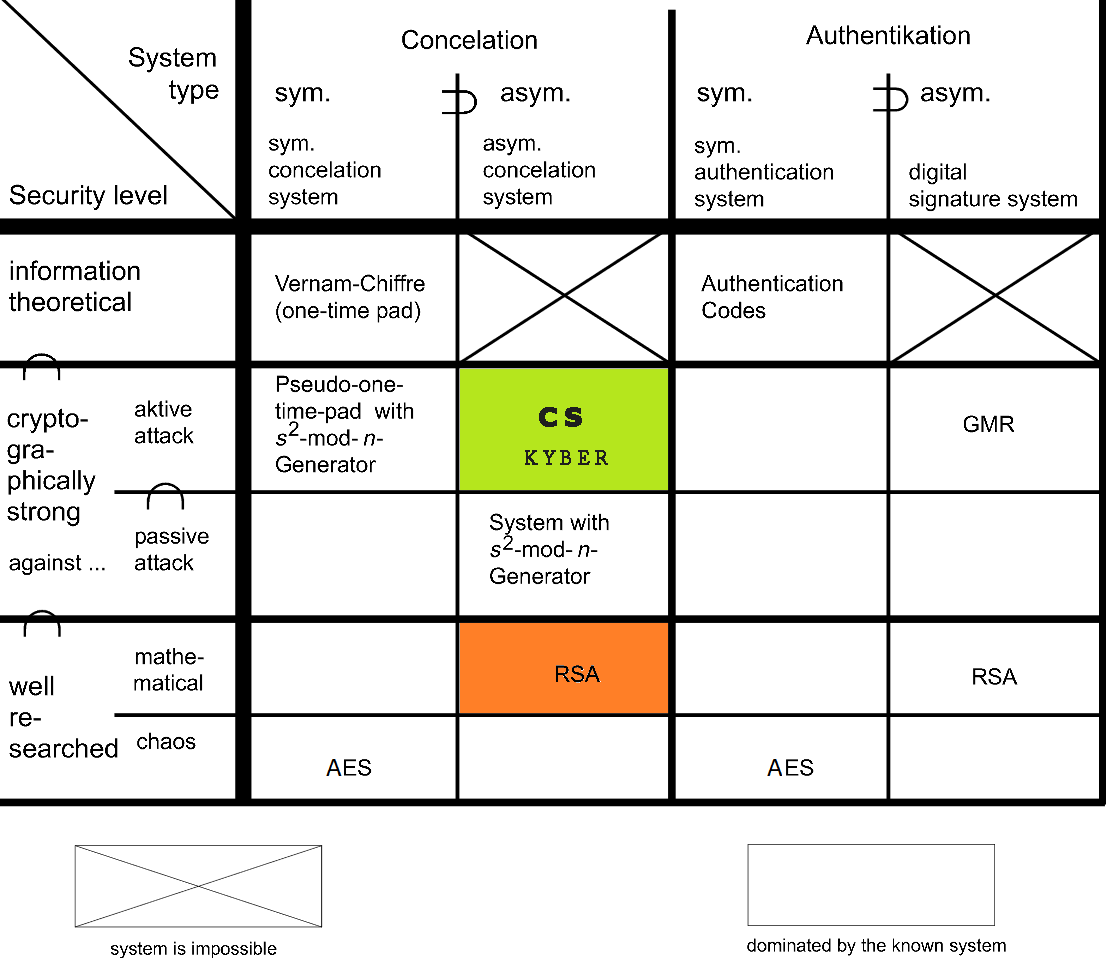}}
	\caption{Overview on cryptographical systems in relation to security level~\cite{Pfitzmann2006}.}
	\label{fig:OverviewCryptographicSystems}
\end{figure}

Advances in quantum computing have created a need for new methods in PQC.
Over the past years, different cryptography systems have been developed.
%These are basically divided in encryption, also called concelation, and signature systems.
%Each category is further distinguished in public-key, also called asymmetric, and symmetric systems.
Many schema can be used and combined for multiple security operations. % like encryption, also called concelation, and signing.
An overview about established cryptographic systems and their security level is given in Figure \ref{fig:OverviewCryptographicSystems}.
%well-regarded asymmetric key techniques
It shows the theoretical limits for their level of security in the different categories.
As this publication focus on public-key encryption, the highest security level can only be cryptographic strong. %according to which
A public-key crypto system can never be information theoretical secure due to the public-key testing possibility.
%\todo[inline]{Bild vervollständigen}
%
The following list of encryption methods illustrates the known security levels based on their historical development.
One of the first public-key crypto systems was developed by Ralph Merkle with the Merkle Puzzle, later published in 1978 \cite{Merkle1978}.
Nevertheless, James H. Ellis, Clifford Cocks, and Malcolm Williamson invented public-key cryptography for the British Government Communications Headquarters in 1970 \cite{COMMUNICATIONSELECTRONICSSECURITYGROUP1970}.
The first wide spread protocol for asymmetric cryptography is the Diffie–Hellman(-Merkle)~(DH) key exchange for a non-authenticated key-agreement, since 1976~\cite{Diffie1976}.
Beyond that, the first public-key scheme was developed by Rivest, Shamir und Adleman~(RSA) at the Massachusetts Institute of Technology in 1977~\cite{Rivest1978}.
%The RSA crypto system works like a trap door permutation and can be used as encryption and signature.
These are based on the assumption of the hardness of the factoring problem or discrete logarithm problem~(DLP).
Since these systems work deterministic, it is susceptible to simple attacks.
Therefore, for example RSA has to be combined with OAEP in practice nowadays~\cite{Bellare1994}.
The unmodified RSA is not \textit{indistinguishable for chosen plain-text attacks} (IND-CPA), which is mandatory for current systems.
%Nevertheless, RSA with OAEP is not IND-CCA2 secure \cite{Paillier2006}. % Es gibt Möglichkeiten, um dies zu erreichen: Victor Shoup: OAEP Reconsidered. In: CRYPTO 2001
Beside these, there are many more Public-Key schemes with the following security problems (excerpt):
\begin{itemize}
	\item Merkle-Hellman~\cite{1055927}: Trapdoor knapsacks; broken~\cite{1056964}. % (1978) 
	\item McEliece~\cite{McEliece1978}: Provable hardness of decoding a general linear code (IND-CPA); proposed PQC; comparable large keys. %(1978) 
	\item Rabin~\cite{Rabin1979}: Provable for factoring problem; 4-to-1 output, which leads to decryption failure. % Category: well researched; %(1979) 
	\item Chor-Rivest~\cite{Chor1984}: no feasible attack known; comparable large key. %(1984) 
	\item Elgamal~\cite{ElGamal1985}: Provable hardness of decisional Diffie–Hellman assumption (IND-CPA); malleable. % and not IND-CCAcomputational and decisional Diffie–Hellman assumption, random oracle model %(1985) 
	\item NTRUEncrypt~\cite{Hoffstein1998}: Provable hardness for correctness \textit{Ring Learning with Errors}; proposed PQC; comparable large keys. %Assumption: shortest vector problem, factoring certain polynomials; proof on R-LWEproblem; Category: well researched %(1996) 
	%\item Okamoto–Uchiyama (1998) \cite{Okamoto1998}: Provable hardness of factoring problem with subgroups (IND-CPA); malleable. %Assumption: factoring problem/p-subgroup (IND-CPA); malleable
	%\item Naccache–Stern (1998) \cite{Naccache1998}: Provable hardness of quadratic residuosity problem (IND-CPA); malleable. %Assumption: quadratic residuosity problem (IND-CPA); malleable 
	\item Paillier~\cite{Paillier1999}: Provable hardness of decisional composite residuosity problem (IND-CPA); malleable. %Proovable: decisional composite residuosity problem (IND-CPA); malleable% and not IND-CCA2 (in standard form) % (1999) 
	%\item Damgård–Jurik (2001) \cite{Damgaard2001}: Provable hardness of decisional composite residuosity problem (IND-CPA); malleable. %Proovable: decisional composite residuosity problem (IND-CPA); malleable
\end{itemize}

%https://www.nist.gov/news-events/news/2019/01/nist-reveals-26-algorithms-advancing-post-quantum-crypto-semifinals
%https://www.nist.gov/news-events/news/2022/07/nist-announces-first-four-quantum-resistant-cryptographic-algorithms
American NIST and European ENISA are currently running a competition for new methods for quantum-resistant public-key cryptographic~\cite{NIST2022,CybersecurityENISA2022,Alagic2022}. %NIST2022~\cite{Alagic2022}
CRYSTALS-Kyber has been nominated as a possible finalist~\cite{Regev2005,Bos2018}.
Contrary to our EC or Isogeny approach, Kyber is based on the hardness of solving the learning-with-errors (LWE) problem over module lattices.
Thus, the keys are slightly larger with comparable security and a side-channel attack is known~\cite{cryptoeprint:2022/1713}.
Besides the ring-LWE and lattice-based approach, further promising approaches for POC are considered: multivariant polynomial, code-based, hash-based, and supersingular isogeny EC.
Of these, the EC approach offers the most promising potential in terms of lightweight Cryptography.
The key length in EC and supersingular isogeny EC cryptography is significantly smaller than in the other cryptosystems. % based on latticies, hashes, or codes.

%Retrospective
In the following, a retrospective of the development of supersingular isogeny EC cryptography is given:
\begin{itemize}
\item 1996 Couveignes~\cite{cryptoeprint:2006/291} first mentioned about isogenies in cryptography.% but only published in 2006; %1996
%\item %2009  Charles et al presented hash functions constructions based on isogenies;
\item 2010 Rostovtsev and Stolbunov~\cite{Rostovtsev2006} presented first published isogeny-based public-key cryptosystem based on isogenies between ordinary curves. %; %2010  
%\item %2010  Childs et al.  presented a quantum subexponenal attack on Stolbunovs public-key cryptosystem;
\item 2011 Jao and De Feo~\cite{Feo2011} presented the Supersingular Isogeny Diffie-Hellman (SIDH)~\cite{Castryck2022}. % and a identification protocol. %2011  
\item 2017 Jao et al.~\cite{Azarderakhsh2018} proposed Supersingular Isogeny Key Encapsulaon (SIKE) as a submission to NIST PQC call. %2017  
%\item %2018  Castryck et al.  proposed a (Comutative)SIDH based on Couveignes construction
\end{itemize}
The SIDH and SIKE approaches are the only known approaches based on supersingular isogeny.
However, an active attack was found for SIDH~\cite{journals/iacr/GalbraithPST16,Castryck2022}. %cryptoeprint:2022/975
This type of adaptive attacks is fundamentally prevented with our concept.
Furthermore, the NIST calls for \textit{Lightweight Cryptography} to protect small electronics, we are focus on~\cite{NIST2018}.\\

%Due to the developlment on large number factorization, more complex mathematical problems have to be identified for cryptography.
%The direction of elliptic curve cryptography has yielded algorithms based on the discrete logarithm problem by Neal Koblitz and Victor Miller. %1985
%It provides smaller key sizes and faster operations for approximately equivalent estimated security. 
\section{Concept of Cramer-Shoup with Elliptic Curve}
%this approach is based on the general sender-receiver architecture. 
This approach is premised on the general sender-receiver architecture, as shown in Figure~\ref{fig:scenario}. 
Our public-key encryption method is based on the approach of Cramer-Shoup (CS)~\cite{Cramer1998,cryptoeprint:2003/087}.
Here, we adapt the cryptographic strong procedure of CS to the promising base of ECC as an intermediate step to supersingular isogeny EC.
The main benefit is that the key length scales linear in relation to the security level~\cite{Giry2023}.
The new security relies on the \textit{Elliptic Curve Discrete Logarithm Problem} (ECDLP)~\cite{Miller1986,Koblitz1987}.
This mathematical problem is more difficult to solve than the integer factorization problem or the classical DLP.
Currently, no algorithm is known that solves the ECDLP in an efficient way.
The known fastest approach is the parallelized Pollard-Rho Algorithms with approximately $O({\sqrt {p}})$, where $p$ is the largest prime factor of $n$~\cite{Seet2007}.
%Due to this, the key length in EC and supersingular isogeny EC cryptography is significantly smaller than in other cryptosystems based on latticies, hashes, or codes.

\subsection{Prerequisite}
First of all, the following parameters are to be declared.
The plain-text message to be secured, is described as the parameter~$m$. % with length~$l$.
Here it is a positive integer value, represented as binary. % that specifies the field $F_{2^m}$.
%It belongs to the integer group $Z$, representing the amount of possible input data.

%We declare $Z$ as an integer group, representing 
%The input messages to be secured is declared as integer group $Z$..

For encryption, we chose two large prime numbers $p$ and $q$ secretly, where as $p = 1 + 2q$.
This defines us the integer group $Z$ over $p$ and $q$, called $Z_p$ and $Z_q$.
The group $G$ is defined as a subgroup of $Z_p$ of order $q$.
A plain-text input $m$ need to be part of the Group $G$, representing the amount of possible input data..
Larger information need to be divided into chunks, so that $m \in G$.

Furthermore, publicly available is a hash function $Hash$, which is a collision resistant one-way function in $Z$. %$Z_q$.
This hash function $Hash$ calculates for any input values in $Z$ integer values as output.
We suggest the standardized SHA3, the sponge function Keccak~\cite{Bertoni2011} for hardware or the more Side-Channel-Attacks robust package Skein~\cite{Ferguson2010} for software.
%H = SHA3 (Keccak) for Hardware or Skein (more robust against Side-Channel-Attacks) for Software.

We use an EC $F_p(x)$ in the finite body modulo $Z_p$ as the basis of our encryption system, whereas $x$ is the input parameter to $F_p$.
More specific, $x$ is an integer coordinate of a point $P(x,y)$ in Cartesian coordinate system.
We suggest an EC $F_p(x)$ fulfilling the Weierstrass form: $y^2 = x^3 + ax + b$ % (Weierstrass form)
The factors $a$, $b \in Z_p$ specifies the field $F_p$, whereas $4a^3 + 27b^2 \; != 0$.

%------------
%The factors $a$ and $b$ are elements of $Z_p$ that specifies the field $F_p$, whereas $4a^3+27b^2 != 0$.
%define the curve E %t
For this purpose, the well-reviewed Koblitz curve \textit{SECP256k1}~\cite{Brown2010} or Montgomery curve \textit{ed25519} (RFC7748)~\cite{Langley2016} are suitable~\cite{Bernstein2013} for a \~128-bit security level.
%\todo[inline]{hier noch beschreiben, dass Punkte auf der Kurve mit Grossbuchstaben bezeichnet werden; Die Funktionen pointAdd und Multi drei parameter haben}
%In the following, points on the curve are defined in capital letters.
Each EC comes with his own public starting-points, which has been identified as cryptographically fitting.
These starting-points are generators of a large cyclic subgroup of the specific curve.
For these points applies $G_1, G_2 \in F_p(x)$.
Currently, our two starting points $G_1(x_{g1},y_{g1})$ and $G_2(x_{g2},y_{g2})$ are chosen wisely random~\cite{Roy2014} in a large cyclic group.

For cryptographic operations, the following functions are described on an EC.
%For the addition of points on the elliptic curve, t
The addition functions is defined with $pointAdd[]$ with the three parameter: the EC $F_p(x)$ itself, point $P_1$, and point $P_2$, so $F_p(P_1 + P_2) = pointAdd[F_p(x), P_1, P_2]$.
The multiplication is described with $pointMult[]$ with the three parameter: the EC $F_p(x)$ itself, starting point $P$, and multiplication factor $k$, so $F_p(P_k) = pointMult[F_p(x), P, k]$.
Beside these functions, we need the point conversion $pointNegate[]$, which invert the position of the point by changing the sign of the y-value.

\subsection{Public Key Generation by Receiver}
At the beginning, the receiver chooses the following five factors randomly, each $\in Z_q$: $x_1$, $x_2$, $y_1$, $y_2$, $z$.
%Each number should be about the same large size.
Each number should be large, favored about the same size.
To create the public-key, the receiver calculates the following values, see Equation~\ref{eccsC}, \ref{eccsD}, and \ref{eccsH}:
\begin{equation}
	\label{eccsC}
	\begin{aligned}
	Point~ C ={} & F_p(x_1 \; G_1 \; + \; x_2 \; G_2) = \\
	& pointAdd[ F_p(x),\\ 
	& pointMult[F_p(x), G_1, x_1],\\
	& pointMult[F_p(x), G_2, x_2] ]
	\end{aligned}
\end{equation}
\begin{equation}
	\label{eccsD}
	\begin{aligned}
	Point~ D ={} & F_p(y_1 \; G_1 \; + \; y_2 \; G_2) = \\
	& pointAdd[ F_p(x),\\
	& pointMult[F_p(x), G_1, y_1],\\
	& pointMult[F_p(x), G_2, y_2] ]
	\end{aligned}
\end{equation}
\begin{equation}
	\label{eccsH}
	\begin{aligned}
	Point~ H ={} & F_p(z \; G1) = \\
	& pointMult[F_p(x), G_1, z]
	\end{aligned}
\vspace{0.3cm}
\end{equation}

In summary, we obtain the following individual keys:
\begin{itemize}
\item Public-key \;- Points:~  $C$, $D$, $H$
\item Private-key - Factors: $x_1$, $x_2$, $y_1$, $y_2$, $z$\\
\end{itemize}

In addition, we have in common the following parameters, which can also be public:
\begin{itemize}
\item Function $F_p(X)$ with generator points $G_1$, $G_2$
\item Function $Hash$\\
\end{itemize}

For the public format, recommended representation is ANSI X.509, X9.62, and X9.63 syntax following ASN.1 structure.

%\todo[inline]{check if mod is necessary}
%ja, ist oben allgemein zu F(x) definiert
\subsection{Encryption by Sender}
The sender would like to store or transmit the data $m$. %positive integer that specifies the field $F\_2^m$
For encryption, we secretly and randomly choose a multiplication factor $r \in Z_q$.
The factor $r$ is chosen anew for each \mbox{data $m$.}
%The factor $r$ is selected individually for every new  %secret and individual Multiplikation-Factor
Even if $q$ is unknown and therefore also $Z_q$, $r$ should automatically be part of $Z_q$, because $q$ is chosen accordingly large.
This factor $r$ is used to perform point multiplications on the EC as follows, see Equation~\ref{eccsU1}, \ref{eccsU2}, and \ref{eccsE}:
\begin{equation}
	\label{eccsU1}
	Point~ U_1 = F_p(r \; G_1) = pointMult[F_p(x), G_1, r] \;\;\;
\end{equation}
\begin{equation}
	\label{eccsU2}
	Point~ U_2 = F_p(r \; G_2) = pointMult[F_p(x), G_2, r] \;\;\;
\end{equation}
\begin{equation}
	\label{eccsE}
	\begin{aligned}
	Point~ E & ={} F_p(r \; H \; + \; m) =\\
	& pointAdd[F_p(x), pointMult[F_p(x), H, r], m ]
	\end{aligned}
\end{equation}
There we obtain the three points $U_1$, $U_2$, and $E$ of the EC.

%Tamper Protection
To protect against tampering and to ensure integrity, one hash value $\alpha$ is calculated over the three points, see Equation~\ref{eccshashcalc}:
\begin{equation}
	\label{eccshashcalc}
	\alpha{} = H(U_1, U_2, E)% \; mod \; F_p(x)
\end{equation}
This hash value must also be encrypted before transmission, see Equation~\ref{eccsV}:
\begin{equation}
	\label{eccsV}
	\begin{aligned}
	Point~ V_{enc} ={} & F_p(r \; C + r \; \alpha{} \; D) = \\
	& pointAdd[ F_p(x),\\
	& pointMult[F_p(x), C, r],\\
	& pointMult[F_p(x), D, r \times \alpha ] \; ]
	\end{aligned}
\end{equation}

%Transmission
The encrypted data $enc\{m\}$ for transmission consists of the following components, see Equation~\ref{eccsENC}:
\begin{equation}
	\label{eccsENC}
	enc\{ m \} = \{U_1, U_2, E, V\}
\end{equation}

\subsection{Decryption by Receiver}
The recipient first verifies the integrity of the received message.
For this purpose, we calculate alpha again and compare it with the encrypted version, see Equation~\ref{eccshashcalcverify} and~\ref{eccsVverify}:
\begin{equation}
	\label{eccshashcalcverify}
	\alpha{} = H(U_1, U_2, E)% \; mod \; F_p(x)
\end{equation}
\begin{equation}
	\label{eccsVverify}
	\begin{aligned}
	Point~ V_{dec} = F(V) ={} & F_p(x_1 \; U_1 \; + \; \alpha{} \; y_1 \; U_1 \; + \; x_2 \; U_2 \; + \; \alpha{} \; y_2 \; U_2)\\ 
	pointAdd[ F_p(x),&\\
	pointAdd[ F_p(x),& pointMult[F_p(x), U_1, x_1],\\
	& 				  pointMult[F_p(x),\\
	&				  pointMult[F_p(x), U_1, y_1], \alpha{}] \; ]\\
	pointAdd[ F_p(x),& pointMult[F_p(x), U_2, x_2],\\
	&				  pointMult[F_p(x),\\
	&				  pointMult[F_p(x), U_2, y_2], \alpha{}] \; ]\\
	&				  ]
%	& pointMult[F_p(x), U1, x1 + \alpha{} y1],\\
%	& pointMult[F_p(x), U2, x2 + \alpha{} y2] ]
	\end{aligned}
\end{equation}
If $V_{enc}$ and $V_{dec}$ matches, the received message is unaltered and belongs to the same random number $r$. %can be decoded.

For the decryption of the message, the factor $r$ is extracted from the two points $U_1$ and $E$ and the factor $z$ is indirectly extracted from the point $H$, see Equation~\ref{eccsDEC}.
\begin{equation}
	\label{eccsDEC}
	\begin{aligned}
	dec\{m\} ={} & F_p(E - z \; U_1) \\
	& pointAdd[ F_p(x), E, \\
	& pointNegate[ pointMult[F_p(x), U_1, z] ] ]
	\end{aligned}
\end{equation}

\section{Evaluation}
In the following, the correct operation of the approach is first demonstrated.
The subsequent performance comparison puts our approach in relation to comparable systems.
\subsection{Proof of Correctness}
For a better comprehensibility and compact notation, the description is done without the continuous specification of the elliptic curve $F(x)$.
The correctness of our system is given by: % (based on $F(x)$):
%$U1^{x1+y1\alpha} \; U2^{x2+y2\alpha} = U1^{x1} \; U2^{x2} \; U1^{y1\alpha} \; U2^{y2\alpha} = G1^{r x1} \; G2^{r x2} \; G1^{r y1 \alpha} \;  G2^{r y2 \alpha} = (G1^{x1} G2^{x2})^r \; (G1^{y1} G2^{y2})^{r\alpha} = C^{r} \; D{r \alpha} = V$
\begin{equation}
	%\label{x}
	\begin{aligned}
&\quad (x_1 + y_1 \; \alpha)\; U_1 + (x_2 + y_2 \; \alpha) \; U_2\\
&= x_1 \; U_1 + x_2 \; U_2 + y_1 \; \alpha{} \; U_1 + y_2 \; \alpha{} \; U_2\\
&= r \; x_1 \; G_1 + r \; x_2 \; G_2 + r \; y_1 \; \alpha{} \; G_1 + r \; y_2 \; \alpha{} \; G_2\\
&= r \; (x_1 \; G_1 + x_2 \; G_2) + r \; \alpha{} \; (y_1 \; G_1 + y_2 \; G_2)\\
&= r \; C + r \; \alpha{} \; D = V.\\
	\end{aligned}
\end{equation}

%Since u1^z = h^r , Dec sk ( u 1 ,u 2 ,e,v ) = e/u1^z = e/h r = m.
Since $z \; U_1 = r \; H$, $dec\{m\} = \{U_1,U_2,E,V\}$
\begin{equation}
	%\label{x}
	\begin{aligned}
\;\; = E \times (- z \; U_1) = E \times (- r \; H) = m.
 	\end{aligned}
\end{equation}

So, our system is working properly.

\subsection{Preliminary Performance Comparision}
The implementation of our schema is done from an academic perspective in Java without any special optimizations. %in relation to the performance
It serves to verify and validate the correct functioning.
The performance of the algorithms is highly dependent on proper implementation and mathematical realization.
The Table~\ref{fig:performance} shows a preliminary comparison of the speeds of different encryption algorithms on an Intel Core i7-8565U CPU \@ 1.80GHz.
We used freely available sources as reference implementation for RSA with Chinese-Reminder-Theorem~\footnote{https://github.com/YYZ/RSA}, ECDH~\footnote{http://www.academicpub.org/PaperInfo.aspx?PaperID=14496}, SIDH~\footnote{https://github.com/Art3misOne/sidh}, and Kyber~\footnote{https://github.com/fisherstevenk/kyberJCE}.
%The comparison is made for all procedures with a 256 bit key.
%This would no longer be sufficient for RSA today, but allows a direct relation.
%We calculate the average out of 100 iterations.
%The time for encryption and decryption is given without the preparation for the curve and key generation.
The time is measured in milliseconds.

%\begin{figure}[htbp]
%	\centering
%	\includegraphics[width=1.0\linewidth]{figure/performance.png}
%	%\includegraphics[width=1.0\linewidth]{figure/scenario_general_secrecy_system_shannon.png}
%	\caption[]{Performance comparison of our approach against common reference approaches.}
%	\label{fig:performance}
%\end{figure}

\begin{table}[]
	\centering
	\caption{Comparison of our approach with common reference approaches.}
%	\addtolength{\tabcolsep}{-3pt}
	\addtolength{\tabcolsep}{7pt}
	\footnotesize
	\begin{tabular}{l|cccccccc}
		&\multicolumn{1}{p{0.6cm}}{RSA\newline(4096)}&\multicolumn{1}{p{0.6cm}}{ECC\newline(256)}&\multicolumn{1}{p{0.6cm}}{CS\newline(256)}&\multicolumn{1}{p{0.6cm}}{\textit{THIS}\newline\textit{(256)}}& \multicolumn{1}{p{1.3cm}}{ECDH\newline(secp256k1)}& \multicolumn{1}{p{0.6cm}}{SIDH\newline(P751)}& \multicolumn{1}{p{0.6cm}}{Kyber\newline(1024)}\\
		\hline
		Size Pub. K.              & 512 B                      & 32 B                   & 1 KB                                                     & \textit{64 B}                   & 32 B                                                             & 564 B                   & 1.5 KB                  \\
		Size Pri. K.             & 512 B                      & 32 B                   & 1 KB                                                     & \textit{64 B}                   & 32 B                                                             & 48 B                    & 3.1 KB    \\
		\hline
		Agreement                &      \multicolumn{1}{r}{-}                  &  \multicolumn{1}{r}{-}                &                        \multicolumn{1}{r}{-}              &    \multicolumn{1}{r}{-}           & \multicolumn{1}{c}{2}                                          & \multicolumn{1}{c}{416} &     \multicolumn{1}{r}{-}            \\
		Encryption                   & \multicolumn{1}{r}{116}    & \multicolumn{1}{r}{19} & \multicolumn{1}{r}{3}                                  & \multicolumn{1}{r}{\textit{41}} &                                 \multicolumn{1}{c}{-}                               &   \multicolumn{1}{c}{-}           & \multicolumn{1}{r}{2} \\
		Decryption                   & \multicolumn{1}{r}{4}      & \multicolumn{1}{r}{7}  & \multicolumn{1}{r}{1}                                  & \multicolumn{1}{r}{\textit{43}} &                  \multicolumn{1}{c}{-}                                             &            \multicolumn{1}{c}{-}          & \multicolumn{1}{r}{4} \\
		Initialisation & \multicolumn{1}{r}{17.700} &  \multicolumn{1}{r}{397} & \multicolumn{1}{r}{3}                                    & \multicolumn{1}{r}{\textit{473}} & \multicolumn{1}{c}{682}                                          & \multicolumn{1}{c}{687} & \multicolumn{1}{r}{152} \\
		
	\end{tabular}
	\label{fig:performance}
\end{table}

The ECDH and SIDH~\cite{Costello2016} method require more computing power than our approach and the key cannot be reused.
The speed of RSA depends heavily on the key size, especially the key generation.
However, RSA requires further power through OAEP and key validation, which is not included here.
The more complex basis of our schema requires correspondingly more computing power to guarantee the desired security.
Only the optimized implementation of Kyber for the NIST competition is faster, requiring larger keys~\footnote{https://pq-crystals.org/kyber/}.

%Additionally, our approach can be made polymorphic in the sense of a variable usage of the underlying EC and the flexible choice of the starting points.
%This further complicates a cryptographic analysis and enlarges the possible space of cryptograms.

\section{Proof: Secure against adaptive-choosen ciphertext attacks}
Our presented crypto schema is cryptographic strong, so we can proven the resistance against CCA.
The evidence for CPA is therefore obsolete.
In short, even without having to get too deep into the proof, we refer to existing once for the fundamental CS schema~\cite{Cramer2003,Hastad2003,Paillier2006}.
However, against ECC have been identified some theoretical attack approaches~\cite{GALBRAITH2016}.
A part of them use the currently strongest attack vector based on active attacks, which is directly countered by our schema.

The proof on security is given by contradiction based on the EC $F(x)$ and follows~\cite{Hastad2003,Chen2014}.
The main advantage of the proof is that it does not relay on a zero-knowledge assumption.
%For a better comprehensibility and compact notation, the description is done without the continuous specification of the elliptic curve $F(x)$.
\subsection{DDH Assumption}
%It is based on the Distinguish Diffie-Hellman (DDH) triples ${g^a, g^b, g^{ab}}$ with random $a$, and $b$ from non-Diffie-Hellman triples ${g^a, g^b, g^{c}}$, where $a$, $b$, and $c$ are independent.
The security is based on the mathematical problem of the Decisional Diffie-Hellman~(DDH) triples as computational hardness assumption.
This means that the triples $\{g^a, g^b, g^{ab}\}$ with random $a$, and $b$ are independent from non-Diffie-Hellman triples $\{g^a, g^b, g^{c}\}$, where $a$, $b$, and $c$.
%It is not to distinguish and it is not possible to efficiently compute discrete logs in the multiplicative cyclic group $G$ of order $q$ with generator $g$,.
In the multiplicative cyclic group $G$ of order $q$ with generator $g$, discrete logarithms are indistinguishable and cannot be computed efficiently.

\subsection{CCA Assumption}
We assume a decryption "oracle" that correctly decrypts any given ciphertext.
An attacker chooses two messages $m_1$ and $m_2$, where $m_1 \neq m_2$.
These both messages are send to an encryption service, which only returns randomly one of the messages encrypted.
The attacker is allowed a polynomial-time access to our decryption "oracle", also after obtaining a ciphertext returned from the encryption service.
The direct transmission of a ciphertext is excluded in this case.
The attacker now guesses which message the encryption service has provided.
If this fits better with a probability than \mbox{1/2 + $\delta$}, then the opponent has an advantage defined by $\delta$.

A crypto system is said to be indistinguishable chosen ciphertext attacks (IND-CCA) secure, if the advantage $\delta$ is negligible for any polynomial time attacker.

%Assume that there is a “decryption oracle” that correctly decrypts any given ciphertext.
%An adversary chooses two unequal messages, m 0, m 1 , gives them both to an “encryption oracle” which returns one of the messages encrypted, but the adversary does not know which one.
%The adversary is allowed continued (polynomial-time) access to the “decryption oracle”, even after receiving the ciphertext returned from the encryption oracle, but may not directly submit this particular ciphertext.
%If the adversary can do better than merely guess which message the encryption oracle encrypted, being correct with a probability of 1/2 + e , then the adversary has an advantage defined to be e.
%A cryptosystem is said to by CCA-2 secure if advantage e is negligible for any polynomial time adversary.

\subsection{IND-CCA 1 - non-adaptive Security}
From the public key, the attacker can get the information: % log d = $y1$ + w $y2$.
\begin{equation}
	%\label{x}
	\begin{aligned}
D = G_1^{\; y_1} + w \; G_1^{\; y_2}
 	\end{aligned}
\end{equation}

For a query $<U_1, U_2, E, V>$, we obtain
\begin{equation}
	%\label{x}
	\begin{aligned}
U_1 = G_1^{\; r_1},\; U_2 = G_2^{\; r_2}, r_1 \neq r_2
 	\end{aligned}
\end{equation}

If it is accepted, then $V = U_1^{\; y_1} \; U_2^{\; y_2}$, i. e. the following:
%log V = $r'1 y1$ + w $r'2 y2$.
\begin{equation}
	%\label{x}
	\begin{aligned}
V = r_1 \; G_1^{\; y_1} + r_2 \; w \; G_1^{\; y_2}
 	\end{aligned}
\end{equation}

Since these equations are linearly independent, this happens with only negligible probability.
Based on the validity check, the cases can be proved and the schema is IND-CCA 1 secure.

\subsection{IND-CCA 2 - adaptive Security (Validity Checking Failure)}
For this proof, we need to divide the value of the secret \mbox{key $z$} in $z_1$ and $z_2$.
From the public key $V$, the attacker can get the following information:
\begin{equation}
	%\label{x}
	\begin{aligned}
H = G_1^{\; z_1} \times w \; G_1^{\; z_2}
 	\end{aligned}
\end{equation}

Suppose that this is not a DDH tuple:
\begin{equation}
	%\label{x}
	\begin{aligned}
D = \{U_1 = G_1^{\; r_1}, \; U_2 = G_2^{\; r_2}, r_1 \neq r_2\} %ungleich
 	\end{aligned}
\end{equation}

Then the challenge ciphertext is as follows:
\begin{equation}
	%\label{x}
	\begin{aligned}
&enc\{m\} = <U_1, U_2, E_{DDH}, V_{dec}> =\\
&<U_1, \; U_2, \; U_1^{\; x_1} \; U_2^{\; x_2} \; m, \; U_1^{\; y_1} \; U_2^{\; y_2} \; U_1^{\; z_1 \alpha} \; U_2^{\; z_2 \alpha}>,\\
&where \; \alpha = Hash\{U_1, \; U_2, \; E\}
 	\end{aligned}
\end{equation}

Therefore, the attacker can get the following information:
\begin{equation}
	%\label{x}
	\begin{aligned}
H = r_1 \; U_1^{\; y_1} + r_2 \; w \; U_2^{\; y_2} + r_1 \; G_1^{\; z_1 \; \alpha} + r_2 \; w \; G_1^{\; z_2 \; \alpha}
 	\end{aligned}
\end{equation}

If the attacker queries an invalid ciphertext to the decryption oracle, say:
\begin{equation}
	%\label{x}
	\begin{aligned}
&<U'_1, \; U'_2, \; E', \; V'>,\\
&where \; U'_1 = G_1^{\; r'_1}, \; U'_2 = G_2^{\;r'_2} \; and \; r'_1 \neq r'_2
 	\end{aligned}
\end{equation}

As for this decryption query, we should consider the followings cases:
\begin{itemize}
\item If $<U_1, \; U_2, \; E> = <U'_1, \; U'_2, \; E'>$ then $V \neq V'$.\\
This query will always be rejected.
\item If $<U_1, \; U_2, \; E> = <U'_1, \; U'_2, \; E'>$ then $V = V'$.\\
Since $Hash$ is collision-resistant and the attack runs in polynomial time, this happens with only negligible probability.
\item If $Hash\{U_1, \; U_2, \; E\} \neq Hash\{U'_1, \; U'_2, \; E'\}$:\\
And if the ciphertext is accepted by the oracle, it should satisfy the following:\\
$H' = r'_1 \; U_1^{\; y_1} + r'_2 \; w \; U_2^{\; y_2} + r'_1 \; G_1^{\; z_1 \; \alpha'} + r'2 \; w \; G_1^{\; z_2 \; \alpha'}$,\\
where $\alpha'$ = $Hash\{U'_1, U'_2, E'\}$.\\
Since the above equations are linearly independent, this happens only with negligible probability.
\end{itemize}
Based on the validity check, all cases can be proved and the schema is IND-CCA 2 secure.
This is currently the strongest notion of security.
In addition, our cryptographic system is highly efficient in terms of computation, especially in the context of hybrid systems for encryption and signature.
%Furthermore, our crypto system is highly computation efficient, especially in relation of hybrid systems for encryption and signation.
In addition, the comparatively small key size enables the system to be used in mobile and wireless applications with low transmission bandwidth, such as smart cards.
This also makes it ideal for the Internet of Things and banking.

%CS is mathematical proof secure against:
%passive, overserving: Interesting in the contents of foreign communication without interferring
%passive, changing: breaks access control and tries to decrypt forgein encrypted files
%active, oberserving: allowed selection of cleartexts, which are then decrypted with a key unkown to the attacker
%active, changing: breaks access control and tries to decrypt encrypted files which enclose cleartext partially choosen
\section{Security discussion: Post-Quantum Cryptography}
Peter Shor developed a polynomial time quantum computer algorithm to solve integer factorization problem and DLP~\cite{PeterShor1994}.
Cryptographic schemes based on pure EC might be not be secure for future, due to the rapid development of quantum technology and data storing possibilities.
What cannot be cracked today can be stored for later decryption~\cite{Burr2013}.
%Anzahl an notwendigen Qbits diskutieren
Currently, a quantum computer needs for breaking an ECC with 256 bit keys (128 bit security level) about 2330 qbits and 126 billion Toffoli gates~\cite{Roetteler2017}.
%This exceeds any current quantum computer approach with less than 400 qbits and seems to be more than a decade in the future.
This exceeds any current quantum computing approach of currently less than 400 Qbits and appears to be more than a decade in the future.
According to NIST and the German BSI, a key length of 256 bit in ECC provide security beyond the year 2030~\cite{NIST2020,FOIS2023}. %NIST2020
Additionally, our approach can be made polymorphic in the sense of a variable usage of the underlying EC and the flexible choice of the starting points.
This further complicates a cryptographic analysis and enlarges the possible space of cryptograms.

Nevertheless, the adaption of the CS scheme to EC was mandatory beforehand to enhance it to an supersingular isogeny EC base~\cite{Jao2011,Feo2014}.
ECC with Montgomery curves has usually corresponding isomorphic Weierstrass curve over a field K in the form: $F(x): By^2 = x^3 + Ax^2 + x$ ~\cite{Velu1972,Montgomery1987,Biasse2014}. %
This is used to enhance our system to the base of supersingular isogeny EC cryptography in a next step.
%
%We suggest a curve $F(x)$ fulfilling the Weierstrass form: $y^2 = x^3 + ax + b$ % (Weierstrass form)
%The factors $a$, $b \in Z_p$ specifies the field $F_p$, whereas $4a^3 + 27b^2 != 0$.
The security is related to the problem of finding the isogeny mapping between two supersingular EC with the same number of points.
Best known attacks are Meet-in-the-Middle \cite{Tani2009}, collision search \cite{Adj2018}, and algorithmic computation \cite{Biasse2014}.
The security will be $O(p^{1/4})$ for classical computers and $O(p^{1/6})$ for quantum computers.
For a classical security level of 128 bit, we need primes of size at least of 768 bit~\cite{Jao2011,GALBRAITH2016}

These isogeny approaches are promising and based on complex problems, which are also resistant in the post-quantum computing era, like SIDE and SIKE.
Although, these new mathematical construction is not the mainstream research for post-quantum cryptography, it offers promising possibilities.
The key sizes are significantly smaller in relation to other schemes. % based on lattices, code-based, and multivariants.
With key-compression techniques, the transmit information with coefficients defining the EC and two EC points is $<$ 517 Bytes~\cite{Costello2016}.
So this fits easily in the payload of one IPv4 or v6 network packet.
It is especially favorable for smart cards and low bandwidth communication as stated in ISO/IEC 7816-8.

\section{Summary}
Although there are not yet sufficiently powerful quantum computers to break the public-key methods currently in use, this could be the case in the distant future.
%What cannot be cracked today can be stored for later decryption~\cite{Burr2013}.
Therefore, research is already being conducted on secure schemes in many different aspects. %~\cite{Alagic2022}.
Our approach follows the transformation to EC and supersingular isgoeny EC like DH over ECDH to SIDH and SIKE.
This paper adapt and enhances the cryptographic strong procedure of Cramer-Soup to the base of EC.
In relation to other suggested crypto system, we focus on \textit{Lightweight Cryptography}.
The main advantage of our system is the comparable higher security than RSA or other approaches by small key size and linear key scaling.
So, our schema can be used in mobile systems with limited bandwidth or less capacity like smart cards or RFID.
Our public-key encryption schema is provable secure IND-CCA~2 without malleability to prevent attacks like from \textit{Bleichenbacher} from the beginning.
%
%The approach of SIKE is based on isogeny ellpitic curve, but faced some attacks~\cite{Castryck2022}.
%It represents a necessary intermediate step on the way to a PQC resistant method based in supersingular isogeny EC. %and Shores algorithm~\cite{PeterShor1994}
%
In the future, we will adapt our encryption system to supersingular isogeny EC to foster resist quantum computing capabilities.

\bibliographystyle{splncs04}
\bibliography{IEEEexample}

\begin{thebibliography}{10}
\providecommand{\url}[1]{\texttt{#1}}
\providecommand{\urlprefix}{URL }

\bibitem{Bleichenbacher:1998:CCA:646763.706320}
Bleichenbacher, D.: {Chosen Ciphertext Attacks Against Protocols Based on the
  RSA Encryption Standard PKCS \#1}. In: Proceedings of the International
  Cryptology Conference on Advances in Cryptology (CRYPTO). pp. 1--12.
  Springer, London, UK (1998),
  \url{http://dl.acm.org/citation.cfm?id=646763.706320}

\bibitem{Bellare1994}
Bellare, M., Rogaway, P.: {Optimal Asymmetric Encryption How to Encrypt with
  RSA}. Advances in Cryptology - Eurocrypt  (1994)

\bibitem{Fujisaki2004}
Fujisaki, E., Okamoto, T., Pointcheval, D., Stern, J.: {RSA-OAEP Is Secure
  under the RSA Assumption}. Journal of Cryptology  17(2),  81--104 (Mar 2004)

\bibitem{Canetti:2004:ROM:1008731.1008734}
Canetti, R., Goldreich, O., Halevi, S.: The random oracle methodology,
  revisited. J. ACM  51(4),  557--594 (Jul 2004)

\bibitem{Paillier2006}
Paillier, P., Villar, J.L.: {Trading One-Wayness Against Chosen-Ciphertext
  Security in Factoring-Based Encryption}. In: Lai, X., Chen, K. (eds.)
  Advances in Cryptology -- ASIACRYPT. pp. 252--266. Springer Berlin
  Heidelberg, Berlin, Heidelberg (2006)

\bibitem{Brown07whathashes}
Brown, D.R.L.: What hashes make rsa-oaep secure? (2007)

\bibitem{217494}
B{\"o}ck, H., Somorovsky, J., Young, C.: {Return Of
  Bleichenbacher{\textquoteright}s Oracle Threat ({ROBOT})}. In: USENIX
  Security Symposium. pp. 817--849. {USENIX} Association, Baltimore, MD (2018),
  \url{https://www.usenix.org/conference/usenixsecurity18/presentation/bock}

\bibitem{Manger2001}
Manger, J.: {A Chosen Ciphertext Attack on RSA Optimal Asymmetric Encryption
  Padding (OAEP) as Standardized in PKCS \#1 v2.0}. In: International
  Association for Cryptologic Research (IACR), Proceedings of the International
  Cryptology Conference on Advances in Cryptology (CRYPTO). vol. 2139, pp.
  260--274. Springer (2001), lecture Notes in Computer Science

\bibitem{Ronen2018}
Ronen, E., Gillham, R., Genkin, D., Shamir, A., Wong, D., Yarom, Y.: {The 9
  Lives of Bleichenbacher’s CAT:New Cache ATtacks on TLS Implementations}.
  Real World Crypto 2020 and IEEE Symposium on Security and Privacy  (2019)

\bibitem{Heiland2022}
Heiland, E., Hillmann, P.: {(B)LOCKBOX -- Secure Software Architecture with
  Blockchain Verification}. The European Multidisciplinary Society for
  Modelling and Simulation Technology (EUROSIS)  (2022)

\bibitem{Cramer2003}
Cramer, R., Shoup, V.: {Design and Analysis of Practical Public-Key Encryption
  Schemes Secure against Adaptive Chosen Ciphertext Attack}. Aarhus University,
  New York University  (2003)

\bibitem{Hillmann2019}
Hillmann, P., Kn\"upfer, M., Guggemos, T., Streit, K.: {CAKE: An Efficient
  Group Key Management for Dynamic Groups}. INFOCOMP Journal of Computer
  Science  18(2) (2019)

\bibitem{Shannon1946}
Shannon, C.E.: {A Mathematical Theory of Cryptography}. {Communication Theory
  of Secrecy Systems}  (1946)

\bibitem{Bellare1998}
Bellare, M., Desai, A., Pointcheval, D., Rogaway, P.: {Relations among notions
  of security for public-key encryption schemes}. Springer, Advances in
  Cryptology (CRYPTO)  (1998)

\bibitem{Bernstein2018}
Bernstein, D.J.: {The libpqcryptosoftware library forpost-quantum cryptography}
  (2018),
  \url{https://cr.yp.to/talks/2018.05.09/slides-djb-20180509-libpqcrypto-4x3.pdf}

\bibitem{Pfitzmann2006}
Pfitzmann, A.: {Security in IT Networks: Multilateral Security in Distributed
  and by Distributed Systems} (2006)

\bibitem{Merkle1978}
Merkle, R.C.: {Secure Communications Over Insecure Channels}. In:
  Communications of the ACM. 21. pp. 294--299 (1978)

\bibitem{COMMUNICATIONSELECTRONICSSECURITYGROUP1970}
{Communications Electronics Security Group}: {The Possibility of Secure
  Non-Secret Digital Encryption}. Research Report No. 3006  (1970),
  \url{https://www.gchq.gov.uk/sites/default/files/document_files/CESG_Research_Report_No_3006_0.pdf}

\bibitem{Diffie1976}
Diffie, W., Hellmann, M.E.: {New Directions in Cryptography}. IEEE Transactions
  on Information Theory  (1976)

\bibitem{Rivest1978}
Rivest, R.L., Shamir, A., Adleman, L.: {A Method for Obtaining Digital
  Signatures and Public-Key Cryptosystems}. Communications of the ACM  (1978)

\bibitem{1055927}
Merkle, R., Hellman, M.: Hiding information and signatures in trapdoor
  knapsacks. IEEE Transactions on Information Theory  24(5),  525--530 (1978)

\bibitem{1056964}
Shamir, A.: A polynomial-time algorithm for breaking the basic merkle - hellman
  cryptosystem. IEEE Transactions on Information Theory  30(5),  699--704
  (1984)

\bibitem{McEliece1978}
McEliece, R.J.: {A Public-Key Cryptosystem Based on Algebraic Coding Theory}.
  Deep Space Network Progress Report pp. 114--116 (1978)

\bibitem{Rabin1979}
Rabin, M.O.: Digitalized signatures and public-key functions as intractable as
  factorization. MIT-LCS-TR 212, MIT Laboratory for Computer Science  (1979)

\bibitem{Chor1984}
Chor, B., Rivest, R.L.: {A Knapsack Type Public Key CryptosystemBased On
  Arithmetic in FiniteFields}. Advancesin Cryptology: Proceedingsof CRYPTO,
  Springer pp. 54--65 (1984)

\bibitem{ElGamal1985}
ElGamal, T.: {A Public-Key Cryptosystem and a Signature Scheme Based on
  Discrete Logarithms}. IEEE Transactions on Information Theory pp. 469--472
  (1985)

\bibitem{Hoffstein1998}
Hoffstein, J., Pipher, J., Silverman, J.: {NTRU: A Ring-Based Public Key
  Cryptosystem}. International Algorithmic Number Theory Symposium  (1998)

\bibitem{Paillier1999}
Paillier, P.: {Cryptosystems Based on Composite Residuosity} (1999), {École
  Nationale Supérieure des Télécommunications}

\bibitem{NIST2022}
{National Institute of Standards and Technology}: {NIST Announces First Four
  Quantum-Resistant Cryptographic Algorithms}  (2022),
  \url{https://www.nist.gov/news-events/news/2022/07/nist-announces-first-four-quantum-resistant-cryptographic-algorithms}

\bibitem{CybersecurityENISA2022}
{The European Union Agency for Cybersecurity (ENISA)}: {Post-Quantum
  Cryptography - Integration study}  (2022),
  \url{https://www.enisa.europa.eu/publications/post-quantum-cryptography-integration-study/@@download/fullReport}

\bibitem{Alagic2022}
Alagic, G., Apon, D., Cooper, D., Dang, Q., Dang, T., Kelsey, J., Lichtinger,
  J., Miller, C., Moody, D., Peralta, R., Perlner, R., Robinson, A.,
  Smith-Tone, D., Liu, Y.K.: {Status Report on the Third Round of the NIST
  Post-Quantum Cryptography Standardization Process}. NISTIR 8413  (2022)

\bibitem{Regev2005}
Regev, O.: On lattices, learning with errors, random linear codes, and
  cryptography. ACM symposium on Theory of computing (STOC)  (2005)

\bibitem{Bos2018}
Bos, J., Ducas, L., Kiltz, E., Lepoint, T., Lyubashevsky, V., Schanck, J.M.,
  Schwabe, P., Seiler, G., Stehle, D.: {CRYSTALS-Kyber}. IEEE European
  Symposium on Security and Privacy (EuroS\&P)  (2018),
  \url{https://pq-crystals.org/kyber/resources.shtml}

\bibitem{cryptoeprint:2022/1713}
Dubrova, E., Ngo, K., Gärtner, J.: Breaking a fifth-order masked
  implementation of crystals-kyber by copy-paste. Cryptology ePrint Archive,
  Paper 2022/1713 (2022), \url{https://eprint.iacr.org/2022/1713},
  \url{https://eprint.iacr.org/2022/1713}

\bibitem{cryptoeprint:2006/291}
Couveignes, J.M.: Hard homogeneous spaces. Cryptology ePrint Archive, Paper
  2006/291 (2006), \url{https://eprint.iacr.org/2006/291},
  \url{https://eprint.iacr.org/2006/291}

\bibitem{Rostovtsev2006}
Rostovtsev, A., Stolbunov, A.: {Public-Key Cryptosystem based on Isogenies}
  (2006)

\bibitem{Feo2011}
Feo, L.D., Jao, D., Plut, J.: {Towards Quantum-Resistant Cryptosystems from
  Supersingulare Elliptic Curve Isogenies}. PQCrypto, Springer  (2011)

\bibitem{Castryck2022}
Castryck, W., Decru, T.: {An efficient key recovery attack on SIDH}.
  {Cryptology ePrint Archive}  (2022), \url{https://eprint.iacr.org/2022/975},
  \url{https://eprint.iacr.org/2022/975}

\bibitem{Azarderakhsh2018}
Azarderakhsh, R., Koziel, B., Campagna, M., LaMacchia, B., Costello, C., Longa,
  P., Feo, L.D., Naehrig, M., Hess, B., Renes, J., Jalali, R.A., Soukharev, V.,
  Jao, D., Urbanik, D.: {Supersingular Isogeny Key Encapsulation}. NIST
  PQCrypto candidates  (2018),
  \url{https://csrc.nist.gov/CSRC/media/Presentations/SIKE/images-media/SIKE-April2018.pdf}

\bibitem{journals/iacr/GalbraithPST16}
{Steven D. Galbraith}, {Christophe Petit}, {Barak Shani}, {Yan Bo Ti}: On the
  security of supersingular isogeny cryptosystems. IACR Cryptol. ePrint Arch.
  (2016)

\bibitem{NIST2018}
{National Institute of Standards and Technology}: {NIST Issues First Call for
  Lightweight Cryptography to Protect Small Electronics}  (2018),
  \url{https://www.nist.gov/news-events/news/2018/04/nist-issues-first-call-lightweight-cryptography-protect-small-electronics}

\bibitem{Cramer1998}
Cramer, R., Shoup, V.: {A practical public key cryptosystem provably secure
  againstadaptive chosen ciphertext attack}. {Advaces in Cryptology (Crypto),
  LNCS Springer}  1462,  13–25 (1998)

\bibitem{cryptoeprint:2003/087}
Zhu, H.: {A Practical Elliptic Curve Public Key Encryption Scheme Provably
  Secure Against Adaptive Chosen-message Attack}. Cryptology ePrint Archive,
  Paper 2003/087 (2003), \url{https://eprint.iacr.org/2003/087},
  \url{https://eprint.iacr.org/2003/087}

\bibitem{Giry2023}
Giry, D.: {Cryptographic Key Length Recommendation}. BlueKrypt  (2023),
  \url{https://www.keylength.com/en/4/}

\bibitem{Miller1986}
Miller, V.S.: {Use of elliptic curves in cryptography}. Lecture Notes in
  Computer Science  218,  417--426 (1986)

\bibitem{Koblitz1987}
Koblitz, N.: {Elliptic curve cryptosystems}. Mathemathic Computation  48,
  203--209 (1987)

\bibitem{Seet2007}
Seet, M.Z.: {Elliptic Curve Cryptography: Improving the Pollard-Rho Algorithm}.
  Ph.D. thesis (2007)

\bibitem{Bertoni2011}
Bertoni, G., Daemen, J., Peeters, M., Assche, G.V.: {The Keccak reference}
  (2011)

\bibitem{Ferguson2010}
Ferguson, N., Lucks, S., Schneier, B., Whiting, D., Bellare, M., Kohno, T.,
  Callas, J., Walker, J.: {The Skein Hash Function Family}  (2010)

\bibitem{Brown2010}
Brown, D.R.L.: {SEC 2: Recommended Elliptic Curve Domain Parameters}.
  {Standards for efficient Cryptography 2 (SEC 2), Certicom Research}  (2010),
  \url{http://www.secg.org/sec2-v2.pdf}

\bibitem{Langley2016}
Langley, A., Hamburg, M., Turner, S.: {Elliptic Curves for Security (RFC
  7748)}. {Internet Research Task Force (IRTF)}  (2016),
  \url{https://www.ietf.org/rfc/rfc7748.txt}

\bibitem{Bernstein2013}
Bernstein, D.J., Lange, T.: {SafeCurves: choosing safe curves for
  elliptic-curve cryptography}. {Rigidity}  (2013),
  \url{http://safecurves.cr.yp.to/rigid.html}

\bibitem{Roy2014}
Roy, M., Deb, N., Kumar, A.J.: {Point Generation And Base Point Selection In
  ECC: An Overview}. {International Journal of Advanced Research in Computer
  and Communication Engineering (IJARCCE)}  3,  6711--6713 (2014)

\bibitem{Costello2016}
Costello, C., Jao, D., Longa, P., Naehrig, M., Renes, J., Urbanik, D.:
  {Efficient compression of SIDH public keys}. Cryptology ePrint Archive:
  Report 2016/963  (2016)

\bibitem{Hastad2003}
Hastad, J.: {A Provably Secure Public-Key Cryptosystem}. {Seminars in
  Theoretical Computer Science at NADA, KTH}  (2003)

\bibitem{GALBRAITH2016}
D., S., Galbraith, Petit, C., Shani, B., Ti, Y.B.: {On the security of
  supersingular isogeny cryptosystem}. IACR  (2016)

\bibitem{Chen2014}
Chen, R.: {Cramer-Shoup Encryption}. {University of Wollongong}  (2014)

\bibitem{PeterShor1994}
{Peter Wiliston Shor}: {Algorithms for quantum computation: Discrete logarithms
  and factoring}. {Annual Symposium on Foundations of Computer Science, IEEE
  Computer Society Press} pp. 124--134 (1994)

\bibitem{Burr2013}
Burr, T.: {Shhh … NSA's Utah Data Center may be open already } (2013),
  \url{https://archive.sltrib.com/article.php?id=56915018&itype=CMSID}

\bibitem{Roetteler2017}
Roetteler, M., Naehrig, M., Svore, K.M., Lauter, K.: {Quantum resource
  estimates for computing elliptic curve discrete logarithms}. Quantum Physics
  (2017)

\bibitem{NIST2020}
{National Institute of Standards and Technology}: {Recommendation forKey
  Managem}. NIST Special Publication 8  (2020)

\bibitem{FOIS2023}
{Federal Office for Information Security}: {Cryptographic
  Mechanisms:Recommendations and Key Length}. BSI Technical Guide, BSI TR-02102
   (2023),
  \url{https://www.bsi.bund.de/SharedDocs/Downloads/EN/BSI/Publications/TechGuidelines/TG02102/BSI-TR-02102-1.pdf?__blob=publicationFile}

\bibitem{Jao2011}
Jao, D., Feo, L.D.: Towards quantum-resistant cryptosystems from supersingular
  elliptic curve isogenies. In: International Workshop on Post-Quantum
  Cryptography. pp. 19--34 (2011)

\bibitem{Feo2014}
Feo, L.D., Jao, D., Plût, J.: Towards quantum-resistant cryptosystems from
  supersingular elliptic curve isogenies, pp. 209--247 (2014)

\bibitem{Velu1972}
Velu, J.: {Isogenies entre courbes elliptiques}. Comptesrendus de la Academie
  des Sciences  (1971)

\bibitem{Montgomery1987}
Montgomery, P.L.: {Speeding the pollard and elliptic curve methods of
  factorization}. Mathematics of computation  48,  243--264 (1987)

\bibitem{Biasse2014}
Biasse, J.F., Jao, D., Sankar, A.: {A quantum algorithm for computing isogenies
  between supersingular elliptic curves}. CACR  (2014)

\bibitem{Tani2009}
Tani, S.: {Claw finding algorithms using quantum walk}. Theoretical Computer
  Science  (2009)

\bibitem{Adj2018}
Adj, G., Cervantes-Vzquez, D., Chi-Domnguez, J.J., Menezes, A.,
  Rodrguez-Henrquez, F.: {On the cost of computing isogenies between
  supersingular elliptic curves}. Cryptology ePrint Archive, Report 313  (2018)

\end{thebibliography}

%\vspace{2cm}

\section*{Authors}
\noindent {\bf Peter Hillmann} is a postdoctoral researcher and scientific in computer science. He received a M.Sc. in Information-System-Technology from Dresden University of Technology (2011) and a Dr. rer. nat. (Ph.D. in science) degree in Computer Science (2018) from the Universität der Bundeswehr München. He provides expert reports for national and international organizations. His research interests include system and network security with focus on cryptography and IP geolocation as well as middleware technologies and enterprise architecture.\\
%He has been working in the field of cyber security for more than 10 years.\\

%received M.Tech from JNTU Ksd, and he did Bachelor degree in computer science. Currently, he is pursuing his PhD in Computer Science from the University of Ksd. His research interests include Information security, Cryptography in MANET.\\

%\noindent {\bf A.X. Cxz} received Master of computer application from Ksd. He did Bachelor of Science in Information technology from SMU Gang. Currently, he is pursuing his PhD in Computer Science from the University of Jpr. His research interests include Information security, Cryptography, Cyber security, and Algorithms.\\
%
%\noindent {\bf A Qw} received M.Tech from IIT(ISM) Ksd. He did Bachelor degree in Information Technology. Currently, he is pursuing his PhD in Computer Science from the University of Ksd. His research interests include Cryptography, Public key infrastructure.\\

\end{document}